\documentclass{elsart}

\usepackage{amssymb}
\usepackage{graphicx}
\usepackage{bm}
\newcommand{\up}{\uparrow}
\newcommand{\dn}{\downarrow}

\newcommand{\aver}[1]{\langle #1 \rangle}

\newcommand{\diag}{\mathop{\mathrm{diag}}}

\begin{document}

\begin{frontmatter}


\title{Spin current through a tunnel junction}


\author{Matthias Braun$^1$,  J\"urgen K\"onig$^1$, Jan Martinek$^{2,3}$}
\address{$^1$Institut f\"ur Theoretische Physik III, Ruhr-Universit\"at Bochum, 44780 Bochum, Germany}
\address{$^2$Institute of Molecular Physics, Polish Academy of Science, 60-179 Pozna\'n, Poland}
\address{$^3$Institut f\"ur Theoretische Festk\"orperphysik, Universit\"at Karlsruhe, 76128 Karlsruhe, Germany}

\begin{abstract}
We derive an expression for the spin-current through a tunnel barrier in terms 
of many-body Green's functions.
The spin current has two contributions.
One can be associated with angular-momentum transfer by spin-polarized charge 
currents crossing the junction.
If there are magnetic moments on both sides of the tunnel junction, due 
to spin accumulation or ferromagnetic ordering, then there is a second
contribution related to the exchange coupling between the moments.
\end{abstract}


\end{frontmatter}

\section{Introduction}
In recent years spintronics and magneto-electronic 
devices have been a major research topic, biased in part
by the prospect of new technological applications in information 
processing based on the simultaneous use both the electron's spin and
charge degree of freedom \cite{reviews}. 
All spintronic devices depend on driven spin currents between 
different subsystems, which is why a deep understanding of 
these spin-currents is essential for this field of research.

A naive identification of total transfer of angular momentum with 
the spin-polarized charge current passed through a junction,
$I_{\rm spin}=I_\up-I_\dn$, neglects the possibility of spin transfer
due to exchange couplings.
The aim of this paper is to derive an expression for the spin current
through a tunnel junction, that is suitable to describe interacting electron
systems, such as quantum dots, and that explicitly demonstrates the two
different contributions to the spin current.

\section{Derivation of the spin current through a tunnel barrier}
We start our discussion with the calculation of the spin current between a ferromagnetic lead $L$ and an island $I$ via a tunnel contact. In this first section, we do not specify the electronic structure of the island yet, i.e., any kind of many-body effects or couplings of the island to other leads are included. The Hamiltonian of such a tunnel system is given by
\begin{eqnarray}\label{hamiltonian}
H=
\sum_{k\alpha}
\varepsilon^{\,}_{k\alpha}c^{\dag}_{k\alpha}c^{\,}_{k\alpha}
+H_{\rm int}(\{d^\dag_{p\gamma}\};\{d_{p\gamma}\})+\sum_{k\alpha,p\gamma}
\left(V_{k\alpha,p\gamma}c^{\dag}_{k\alpha}d^{}_{p\gamma} + h.c.\right)\,,
\end{eqnarray}
where $c^\dag_{k,\alpha}$ are the fermion
creation operators in the lead and $d^\dag_{p,\gamma}$ 
are the corresponding operator for the island. In 
the lead (island) we label the momentum states of the electrons 
with $k$ ($p$) and the spin with $\alpha$ and $\beta$ ($\gamma$ and $\delta$). 

\begin{figure}[htb]
\begin{center}
\includegraphics[width=0.7\columnwidth,angle=0]{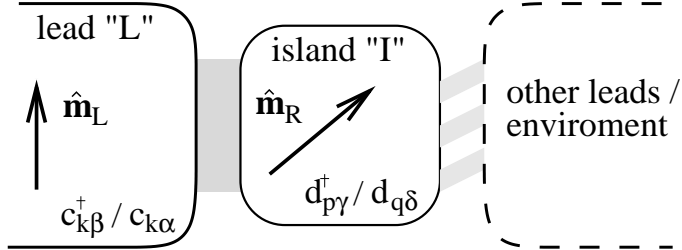}
\end{center}
\caption{A tunnel junction connecting a ferromagnetic 
lead to an island.}
\end{figure}

Starting from this Hamiltonian we calculate the spin 
current in close analogy to the derivation of the 
charge current by Meir and Wingreen for interacting electron systems
\cite{Meir}.
If the spin is a conserved quantity, the time derivative 
of the total spin in the lead equals the spin current 
through the tunnel barrier ${\bm J}_{\rm L} = <\dot{\bm S}_{\rm L}>$.
In the Heisenberg picture, the time evolution of the spin operator
${\bm S}_{\rm L}=(\hbar/2)\sum_{k\alpha\beta} c^{\dag}_{k\alpha} 
{\bm \sigma}^{}_{\alpha\beta} c^{\,}_{k\beta}$ is given by
$i\hbar  \dot{\bm S}_{\rm L}=\left[{\bm S}_{\rm L},H\right]$, which yields

\begin{eqnarray}\label{current1}
  {\bm J}_{\rm L}^{}&=&\frac{-i}{2}
  \sum_{k,p}\left(
  V_{k\alpha,p\gamma}{\bm \sigma}^{\star}_{\alpha\beta}\,\aver{c^\dag_{k\beta}\,d_{p\gamma}} - V_{k\alpha,p\gamma}^\star{\bm \sigma}^{}_{\alpha\beta}\,\aver{c^{}_{k\beta}\,d^{\dag}_{p\gamma}}\right)\\
  &=&\frac{-1}{2}
  \sum_{k,p}\int\frac{d\omega}{2\pi}\left(
V_{k\alpha,p\gamma}{\bm \sigma}^{\star}_{\alpha\beta}\,G^<_{p\gamma,k\beta}(\omega) - V_{k\alpha,p\gamma}^\star{\bm \sigma}^{}_{\alpha\beta}\,G^<_{k\beta,p\gamma}(\omega)\right)\,,\nonumber
\end{eqnarray}
where we introduced the Keldysh Green's functions 
$G^<_{p\gamma,k\beta}(t)=i\aver{c^\dag_{k\beta}(0)\,d_{p\gamma}(t)}$.
By use of a Dyson equation \cite{Meir} we can replace 
the latter with (free) Green's functions 
$g_{k\beta}(\omega)$ of the lead and Green's functions 
$G^<_{q\delta,p\gamma}(t)=i\aver{d^\dag_{p\gamma}\,d_{q\delta}(t)}$ 
of the island. By choosing the magnetization direction 
$\hat{\bf m}_{\rm L}$ as spin-quantization axis, the lead Green`s 
functions $g_{k\sigma}$ are diagonal in the spin index. 
The lead Green's functions are then 
$g^{<}_{k\sigma}= 2\pi i f^+_{\rm L}(\omega)\delta(\omega-\varepsilon_{k\sigma})$, 
$g^{>}_{k\sigma}= -2\pi i f^-_{\rm L}(\omega)\delta(\omega-\varepsilon_{k\sigma})$, 
$g^{\rm ret}_{k\sigma}= 1/(\omega-\varepsilon_{k\sigma}+i0^+)$, 
and $g^{\rm adv}_{k\sigma}=\left(g^{ret}_{k\sigma}\right)^\star$. 
There $f^+_{\rm L}$ stands for the Fermi distribution function 
in the lead $\rm L$ and $f_{\rm L}^-=1-f_{\rm L}^+$.

Assuming, furthermore, that tunnel events conserve the spin 
of the electrons, we substitute the tunnel matrix elements 
by $V_{k\alpha,p\gamma}=t_{k,p}\cdot\delta_{\alpha\gamma}$, 
and define the spin-dependent transition rates 
$\Gamma^\gamma_{q,p}=\sum_{k} t^{}_{k,q}t_{k,p}^\star 
\delta(\omega-\varepsilon_{k\gamma})$. After a lengthy 
but straightforward calculation, the spin current can be written as
\begin{eqnarray}\label{final}
  {\bm J}_{\rm L}^{}
  &=&\!\frac{-2\pi i }{4}\!\!
  \sum_{p,q}\int\frac{d\omega}{2\pi}\,\,\,\,\,
      {\bm\sigma}^{}_{\gamma\delta}(\Gamma^\gamma_{q,p}+\Gamma^\delta_{q,p})\left[ f^+_{\rm L}(\omega)\,G^{>}_{q\delta,p\gamma}+ f^-_{\rm L}(\omega)\,G_{q\delta,p\gamma}^{<}\,\right]\nonumber\\
      &&+{\bm\sigma}^{}_{\gamma\delta}(\Gamma^\gamma_{q,p}-\Gamma^\delta_{q,p})\left[ f^+_{\rm L}(\omega)\,(G^{\rm ret}_{q\delta,p\gamma}+G^{\rm adv}_{q\delta,p\gamma})+\frac{1}{i\pi}{\int}^\prime \!\!dE\,\frac{G_{q\delta,p\gamma}^<(E)}{E-\omega}\,\right]\, .
\end{eqnarray}
This is the central most general result of our calculation. 
Since we did not specify the Green's functions 
$G_{q\delta,p\gamma}$ of the island yet, the expression 
for the spin current can be used for many situation, including
strongly-correlated systems such as quantum dots \cite{ours}.

By comparison with the expression of the charge current derived 
by Meir and Wingreen \cite{Meir} for nonmagnetic systems, we identify
the first line of Eq.~(\ref{final}) with a spin-polarized charge
current.
The origin of the spin-current contribution in the second line 
of Eq.~(\ref{final}) is the exchange interaction between lead and 
island. If the island does posses a magnetic moment, due to 
spin accumulation or ferromagnetic order, this moment couples to the 
magnetization of the lead and both precess around each other \cite{slonprb}. 
This exchange coupling changes the average spin on each 
side of the tunnel junction, and therefore it must also appear 
as contribution to the spin current crossing the tunnel barrier. 
In the latter case, the transfered angular moment is perpendicular to 
the magnetic moments of lead and island, which is sometimes described by
``spin mixing conductances'' \cite{spinmix}.

\section{Application to a FM-FM junction}\label{2}
For concreteness, we restrict the following discussion to the 
special case, that the island is an itinerant ferromagnet. Thereby 
the direction of magnetization $\hat{m}_{\rm I}$ of the island encloses 
a finite angle $\phi$ with the lead magnetization direction.
We further assume, that the island is large, to be also described 
as a reservoir in thermal equilibrium. 
Due to the non-collinear magnetization directions, 
the Green's function of the island is non-diagonal 
in the spin space  
$\check{G}_{p,p}=\check{U}(\phi)\,\diag(g_{p\up},g_{p\dn})\,\check{U}^{-1}(\phi)$
with the SU(2) rotations $\check{U}(\phi)$.

To simplify the result further, we assume that the absolute value of
the tunnel matrix elements $|t^{}_{k,p}|=|t|$ is independent of the 
momentum index. Then we can replace the transition rates 
$\Gamma^\gamma_{q,p}=\sum_{k} t^{}_{k,q}t_{k,p}^\star 
\delta(\omega-\varepsilon_{k,\gamma})$ by the spin-resolved 
density of states
$\rho_{{\rm L},\alpha}=\sum_{k}\delta(\omega-\varepsilon_{k,\alpha})$ and
 $\rho_{{\rm I},\gamma}=\sum_{p}\delta(\omega-\varepsilon_{p,\gamma})$. 
After performing all spin summations, we get in lowest-order in the
tunnel coupling
\begin{eqnarray}\label{FM-FM_end}
{\bm J}_{\rm L}^{}
&=& \frac{\pi}{2} \int\!\! d\omega\,|t|^2 \Biggl(\, 
\Lambda_1(\omega) \hat{\bf m}_{\rm L} + \Lambda_2(\omega)\hat{\bm m}_{\rm I} 
+\Lambda_3(\omega) \hat{\bf m}_{\rm L} \times \hat{\bf m}_{\rm I}\,\Biggl)\\
{\rm with }\, 
&&\Lambda_1(\omega)=[f^-_{\rm L}(\omega)f^+_{\rm I}(\omega)-f^+_{\rm L}(\omega)f^-_{\rm I}(\omega)]\chi_{\rm L}(\omega)\rho_{\rm I}(\omega)\nonumber \\
&&\Lambda_2(\omega)=[f^-_{\rm L}(\omega)f^+_{\rm I}(\omega)-f^+_{\rm L}(\omega)f^-_{\rm I}(\omega)]\rho_{\rm L}(\omega)\chi_{\rm I}(\omega) \nonumber \\
&&\Lambda_3(\omega)=\frac{1}{\pi}{\int}^\prime \!\!dE\,\, \frac{f^+_{\rm L}(\omega)-f^+_{\rm I}(E)}{\omega-E} \chi_{\rm I}(E)\chi_{\rm L}(\omega)\end{eqnarray}
with the full density of states $\rho_{\rm I}=\rho_{\rm I\up}+\rho_{\rm I\dn}$ and the spin-polarization density $\chi_{\rm I}=\rho_{\rm I\up}-\rho_{\rm I\dn}$, and analogue definitions for the lead $L$.

In the first and second term we can recognize the spin 
current contribution of the charge transfer between the 
two reservoirs.
With the cross product $\hat{\bf m}_{\rm I}\times\hat{\bf m}_{\rm L}$  
the third term shows the typical structure 
for a precession movement. In the approach of 
spin dependent circuit theory \cite{spinmix} 
this spin current contribution corresponds to the 
imaginary part of the spin-mixing conductance.

\section{Effect of exchange-coupling contribution}
A very pronounced effect of a spin current is current-induced  
magnetization reversal \cite{mag_reversal}. However, there it is 
difficult to selectively address the spin-current contribution 
arising from the exchange interaction.
The latter goal can be achieved, e.g., in measuring the charge current
through a single-level quantum dot connected to two 
ferromagnetic leads. A current forced 
through such a quantum-dot spin valve will accumulate a 
non-equilibrium spin on the dot. This spin is 
sensitive to the exchange field generated by the leads.
Its precession is predicted to be visible in the magneto-resistance 
of the device \cite{ours}.



\end{document}